\documentclass[10pt, conference, compsocconf, letterpaper]{IEEEtran}
\IEEEoverridecommandlockouts
\usepackage{cite}
\usepackage{amsmath,amssymb,amsfonts}
\usepackage{algorithmic}
\usepackage{graphicx}
\usepackage{textcomp}
\usepackage{subfig}
\usepackage{xcolor}
\usepackage{hyperref}
\usepackage{multirow}
 \def\BibTeX{{\rm B\kern-.05em{\sc i\kern-.025em b}\kern-.08em
    T\kern-.1667em\lower.7ex\hbox{E}\kern-.125emX}}
\begin{document}

\title{OMAD: On-device Mental Anomaly Detection for Substance and Non-Substance Users}
\author{\IEEEauthorblockN{Emon Dey}
\IEEEauthorblockA{Department of Information Systems\\
University of Maryland, Baltimore County (UMBC)\\
Baltimore, USA\\
Email: edey1@umbc.edu}\\\vspace*{-1cm}
\and
\IEEEauthorblockN{Nirmalya Roy}
\IEEEauthorblockA{Department of Information Systems\\
University of Maryland, Baltimore County (UMBC)\\
Baltimore, USA\\
Email: nroy@umbc.edu}\\\vspace*{-1cm}
}

\maketitle

\begin{abstract}
Stay at home order during the COVID-19 helps flatten the curve but ironically, instigate mental health problems among the people who have Substance Use Disorders. Measuring the electrical activity signals in brain using off-the-shelf consumer wearable devices such as smart wristwatch and mapping them in real time to underlying mood, behavioral and emotional changes play striking roles in postulating mental health anomalies. In this work, we propose to implement a wearable, {\it On-device Mental Anomaly Detection (OMAD)} system to detect anomalous behaviors and activities that render to mental health problems and help clinicians to design effective intervention strategies. We propose an intrinsic artifact removal model on Electroencephalogram (EEG) signal to better correlate the fine-grained behavioral changes. We design model compression technique on the artifact removal and activity recognition (main) modules. We implement a magnitude-based weight pruning technique both on convolutional neural network and Multilayer Perceptron to employ the inference phase on Nvidia Jetson Nano; one of the tightest resource-constrained devices for wearables. We experimented with three different combinations of feature extractions and artifact removal approaches. We evaluate the performance of {\it OMAD} in terms of accuracy, F1 score, memory usage and running time for both unpruned and compressed models using EEG data from both control and treatment (alcoholic) groups for different object recognition tasks. Our artifact removal model and main activity detection model achieved about $\approx$ 93\% and 90\% accuracy, respectively with significant reduction in model size (70\%) and inference time (31\%).  
\end{abstract}
\section{Introduction}
Coronavirus disease 2019 (COVID-19) is causing untold challenges to health care providers and other societal entities. Considering the physical health related issues, the older adults of 
65 years and older, people who live in a nursing home or long-term care facility, people with chronic lung, heart, kidney and
liver disease are most at risk. Specially when a respiratory infection is severe, recovery can be prolonged with a general increase in shortness of breath — even after lung function returns to normal \cite{Longterm}. Moreover, present lockdown ordeal is also contributing to another severe health problem: mental disorder, mainly among the vulnerable populations who have a Substance Use Disorders (SUD). Substance Abuse and Mental Health Services Administration (SAMHSA) recognizes there are currently 57.8 million Americans living with substance use disorders (National Survey on Drug Use and Health, 2018) \cite{Samsha}. 

According to a report published in \cite{AARP}, a significant number of COVID-19 positive patients may have irregularities with their nervous system that invokes more intensive research to determine the real implications of this situation. In this respect, evidence of the COVID-19 effects on brain activity is recorded in the MRI of some patients \cite{brain}. Motivated by this, we hypothesize that a deep neural network-integrated edge device that can make decision
on individuals' mental health problems, progressively tracking the brain signal activity can be a useful tool to counter this implication. Deep learning models can enable the Internet of Things (IoT)
devices to interpret unstructured data and intelligently provide feedback to both user and environmental events. However, these works rely on deep networks with millions or even billions of parameters, and the availability of GPUs with very high computation capability \cite{nan2019deep}. This costly requirement of power, memory, and advance computing components may become an impediment to implement deep models on resource constrained devices \cite{tang2017enabling}. 

Researchers have been developing various sophisticated techniques of model compression for deep architectures with negligible compromise of accuracy. For example, a sparse coding-based system to push neural network into the phone's DSP which is an upcoming feature of Google's TensorFlow in partnership with Qualcomm \cite{lane2017squeezing}. The CAE, a DNN-based spike compression model to significantly reduce the data rate of spikes in large-scale neural recording experiments \cite{wu2018lightweight}. If we look into the most recent approaches in the genre, it is conceivable that fundamental theoretical concepts like entropy, game theory, information theory are being incorporated and implemented. The adopted compression and acceleration techniques in these models are sparse coding, orthogonal batch scheduling, segmenting layers into heterogeneous processors etc. The main aim of these methods is to explore the redundancy in the model parameters and remove redundant and uncritical ones \cite{lane2015early} for efficient implementation of signal processing approaches and machine learning models on the embedded and FPGA platforms and devices.

Considering this scope of applying model compression in detecting negative impact of COVID-19 disease from brain activity, we proposed a wearable {\it on-device mental anomaly detection (OMAD)} model to analyze on the EEG data of both substance user and non-substance user. Primarily we have analyzed on liquid substance user (alcohol) on a specific object recognition task. To get a better set of signals for analysis, we have ensured artifact-free dataset by extensive preprocessing steps. We removed the extrinsic type of artifacts using notch filter and designed compressed machine learning models to identify the physiological artifacts. We hypothesize that, an efficient mental behavior anomaly detection approach can alleviate the condition of individuals with co-occurring severe mental illness and substance use disorders, subsequently reducing the rate of substance misuse. To the best our knowledge, it is the first endeavor of incorporating model compression technique with mental anomaly detection application.
The major contributions of this work as follows:
\begin{itemize}
    \item {\it On-device Mental Anomaly detection (OMAD) model:} We propose a wearable device implementable machine learning model to detect mental behavior anomaly for substance and non-substance users. 
    We employ a magnitude-based weight pruning technique to gain a wearable-friendly version of the model. Our compressed model depicts over 90\% accuracy after compression.
    \item {\it Wearable implementable artifact removal model:}
    We design a light-weight artifact removal model to get a smoothed version of the main EEG data. We've collected our own artifact data for model training. This model is used to tag possible artifacts on the main dataset. The same pruning technique is used to reform the original model to a compressed one. We have achieved an approximate 75\% reduced model for artifact removal.
    \item {\it Empirical evaluation on a resource constrained device:} We have used three different experimental settings to demonstrate analysis results. The metrics used are accuracy, F1 score, model size requirement, and inference time for both Multilayer Perceptron (MLP) and CNN based models on Nvidia Jetson Nano, one of the most resource constrained devices available presently. {\it OMAD} achieves approximately 70\% reduction in model size and over 30\% improved inference time with an accuracy loss less than 3\%.
\end{itemize}

\section{Related work}
In this section, we discuss the EEG-based mental health detection approaches and the various model compression-based techniques. We spot the main differences between our approach and those.
\subsection{EEG based mental health detection}
Detecting possible mental disorder among the COVID-19 patients is an ongoing research topic, and researchers are still exploring to find suitable tools and approaches for this purpose. The Electroencephalogram (EEG) based detection system can be efficient because it records the respective brain activity of any action \cite{cohen2017does}. There are numerous amount of works on determining different kinds of mental states using EEG data. Some works related to checking the activity of the paralyzed patients by acquiring their eye movement signals \cite{jiang2019removal}. Also, detecting opioid use disorder using the resting eye state in shown in \cite{wang2015changes}.
Such research works indicates if we can separate the brain signals for similar activities between the control group and a substance user group, it can further be stretched identifying anomalous behavioral activity and can be a useful work for timely intervention. In our work, we have chosen the people who are liquid substance user.

The recorded EEG signal possesses one major data preprocessing challenge due to the various artifacts (EEG noises) in the signal. Different research works worked with different kinds of signal artifact detection and removal. Removing environmental artifacts is comparatively easy
as it can be achieved with selecting filters of different frequency
ranges. However, to remove the extrinsic artifacts, the preprocessing technique has to be more precise \cite{whitton1978spectral}. Highly optimized machine learning models are required for this task to identify those artifacts in
the whole dataset. Works based on eye blinking detection \cite{anderer1999artifact}, muscle movement \cite{rosler2013first}, etc. show different models for efficient detection and removal of those noises to make the main data smoother. In our work, we have collected our own data for eye blinking and eyebrow-raising detection as artifacts.

In our proposed approach, we have combined both the models of artifact detection and mental anomaly detection based on EEG data in our work and also demonstrated results using statistical classifiers, Multilayer Perceptron (MLP) and Convolutional Neural Networks (CNN).
\subsection{Compression techniques}
Developing a deep model based detection is not enough as we have to make the model readily implementable to wearable devices so that real time intervention can be possible. In order to do that we have to try to compress the deep model without compromising much with the accuracy. In recent years researchers have suggested various deep model compression techniques to make this field more enriched. In \cite{han2015learning, li2016pruning}, network pruning and sharing based method is introduced. The main concept behind that is to identify the most contributing nodes. If a node can have an activation function value over a certain threshold, the node will be counted as a valuable node. These selected nodes will be responsible for the output of the future outcome of the model. Another popular approach for compression is quantization and binarization. The core idea is that, in normal computing system, each integer or floating point is represented as 32-bit or 64-bit number. If we can reduce the number of bits required for the representation of the numbers, then eventually, the model size and computing time can be minimized. The authors of \cite{wu2016quantized, gong2014compressing} have represented such quantization-based methods and the main difference between their work, is choice of quantization bit size. They have also implemented their work on some hardwares like Nvidia Titan X \cite{kim2015compression}, Snapdragon 400 \cite{mathur2018using}, 800 \cite{lane2015early} based mobile phones etc.

Moreover, concepts like knowledge distillation \cite{hinton2015distilling}, low rank matrix factorization \cite{zhang2015efficient}, etc. are getting so much traction in recent years. Having the control of changing both the sparsity and scheduling of a deep model persuades us to utilize magnitude-based weight pruning method in our work to squeeze a multilayer perceptron model and an one-dimensional convolutional network with a view to recognizing mental anomaly. The whole squeezed model is implemented on one of the most resource constrained devices named Nvidia Jetson Nano.

\section{Dataset Description}
Our dataset description part is mainly divided into two parts. In the first part we describe the process of data collection for artifact removal and in the next part about the main dataset.
\subsection{Data collection for EEG artifact removal}
EEG plays an important role in identifying brain activity, and behavior. However, the recorded electrical activity always be contaminated with artifacts and then affect the analysis of EEG signal. Artifacts are unwanted signals which are mainly originated from environment noise, experimental error and physiological artifacts. Signal artifacts are more significant while collecting data in EEG form. Furthermore, the environmental noise and experiment error, which come from external factors, are classified as extrinsic artifacts, whereas the physiological signals from body itself (e.g., eye blink, muscle activity, heartbeat) can be categorized as intrinsic artifacts. The extrinsic artifacts can be eliminated by a simple filter due to the frequency of such artifacts are inconsistent with desired signals. Nevertheless, the physiological artifacts are more difficult to be removed as they require particular algorithms.\\
In our data collection, we have considered `eye blinking' and `eyebrow raising' as the EEG artifacts and collected data with {\it Emotiv Epoch EEG headset} with 14 channels. 
The 14 electrodes first have to be soaked with saline water for better connection. When we interface the device with the data collection software, we can see the type of connection between the skull and electrodes. Green signals stands for best connection. 
We have collected 2 trials each from 3 subjects for two kinds of artifacts. Each trial consists of 10 seconds of data, where first 3 and last 3 seconds are resting state and 4 to 7 seconds contain eye blinking and eyebrow raising respectively. The sampling rate chosen is 128 Hz. We have also applied a window size of 128 with 80 percent overlap on the collected dataset.
\subsection{EEG data for substance and non-substance user}
We have used the EEG dataset of UCI machine learning repository for two different groups of control and alcoholic. This data was collected with 64 electrodes sampled at 256 Hz. There were 20 subjects (10 control and 10 alcoholic)in total, and each subject completed 10 trials. Each subject was exposed to either a single stimulus (S1) or to two stimuli (S1 and S2) which were pictures of objects chosen from the 1980 Snodgrass and Vanderwart picture set \cite{snodgrass1980standardized}. A sample dataset is shown in table~\ref{meeg}.
\begin{table}[]
\begin{center}
\caption{Example of main EEG data}
\begin{tabular}{|c|c|c|c|}
\hline
\multicolumn{4}{|l|}{\begin{tabular}[c]{@{}l@{}}\# co2a0000364.rd\\ \# 120 trials, 64 chans, 416 samples\\ \# 3.906000 msecs $\mu$V\\ \# S1 obj, trial 0\\ \# FP1 chan 0\end{tabular}} \\ \hline
0                                         & FP1                                        & 0                                        & -8.921                                        \\ \hline
0                                         & FP1                                        & 1                                        & -8.433                                        \\ \hline
0                                         & FP1                                        & 2                                        & -2.574                                        \\ \hline
0                                         & FP1                                        & 3                                        & 5.239                                         \\ \hline
\end{tabular}
\label{meeg}
\end{center}
\vspace{-4ex}
\end{table}
The first four lines are header information. Line 1 contains the subject identifier and indicates if the subject was an alcoholic (a) or control (c) subject by the fourth letter. Line 4 identifies the matching conditions: a single object shown (S1 obj), object 2 shown in a matching condition (S2 match), and object 2 shown in a non matching condition (S2 nomatch). Line 5 identifies the start of the data from sensor FP1. The four columns of data are: the trial number, sensor position, sample number (0-255), and sensor value (in micro volts).

\section{{\it OMAD} System Architecture}
\begin{figure}[htbp]
\centering
\includegraphics[width=\linewidth]{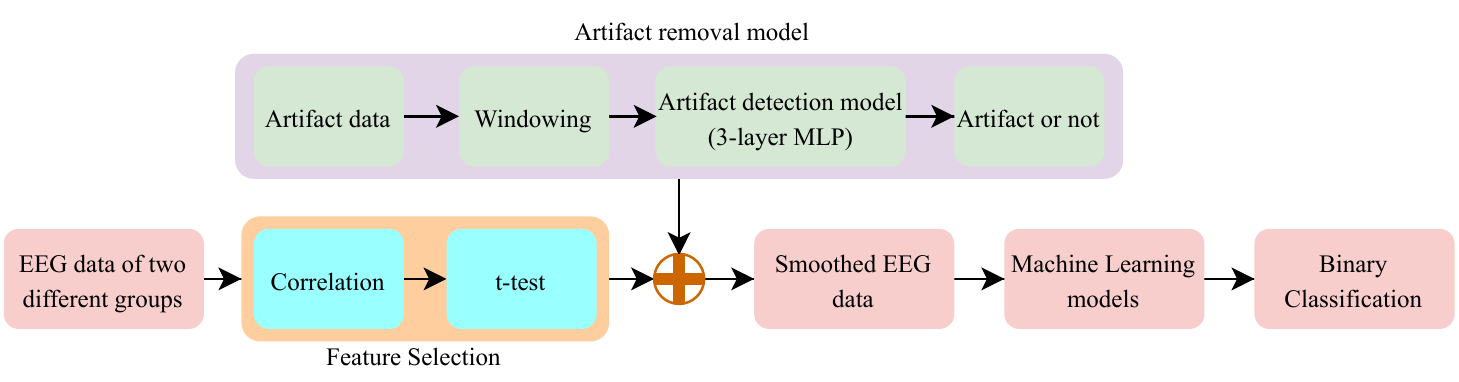}
\caption{Workflow of the {\it OMAD} system.}
\label{overview}
\vspace{-2ex}
\end{figure}

\begin{figure}[htbp]
\centering
\includegraphics[width=\linewidth]{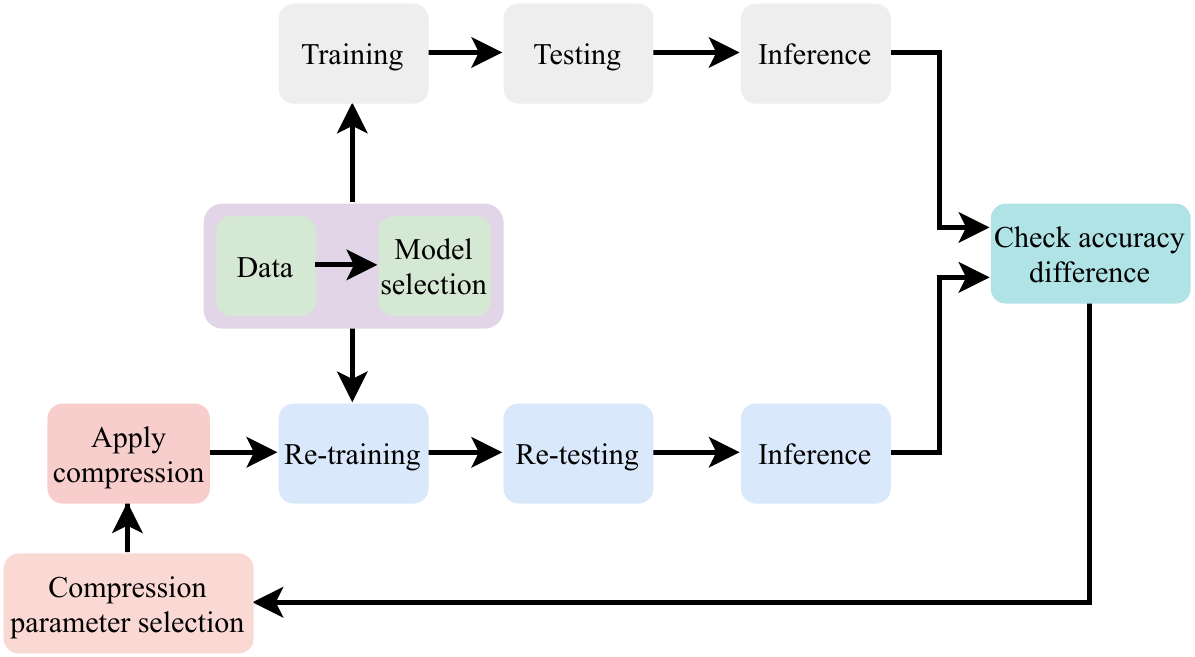}
\caption{Basic functioning procedure of a model compression algorithm.}
\label{comp}
\vspace{-2ex}
\end{figure}

The whole {\it OMAD} system is divided into two parts mainly. First we have collected our own data to develop our artifact removal model. Then the main data are used as the input to the model and further analysis is done based on the output of the removal model. The main EEG data will first go through some feature selection process before going to the artifact removal model. For feature selection we have exploited two common processes named correlation measurement and t-test. The graphical representation is given in fig~\ref{overview}. All of the blocks in the diagram will be described in details in the subsequent sections.
The compression method used here is known as magnitude based weight pruning. The big picture of this process is that, a neural network will be  developed first. Then, compression technique will be applied to the original model and the testing and inference processes are carried out again. The evaluation metrics value for these two different models are compared and if the difference is at satisfactory level, the hyperparameters for this set are 
selected as the resultant one. This process is visualized in fig~\ref{comp}. 
\section{Methodology}
In the following sections we articulate the approach adopted for artifact removal and, model parameter selection. We also include detailed explanation of our compression technique and information of our selected model parameters.
\subsection{Artifact removal procedure}
To analyze the data collected for artifact removal described earlier, we have developed a MLP based model. The only task of the model is to be able to determine which of the data have come from noisy activity. After going through the hyperparameter tuning process, the final model is a 3 layer MLP. Train test ratio is 70:30. The highest test accuracy gained for the primary unpruned model is 95.13\%. Then it is subjected to magnitude based weight pruning procedure and the accuracy after pruning is 92.78\% but the model size is reduced over 75\%. The detailed results of the compressed artifact removal  model is tabulated in table~\ref{artanalysis}. The compression process will be elaborated in later section. The main EEG dataset is then fed to this model and the output of this model will be a binary classification; whether the signal is artifact or not. The signals to which the artifact removal model assigns a tag as artifacts, we'll remove those from the main model. The rest of the signals will then be used for further analysis of classifying the group type.
\begin{table}[htbp]
\caption{Analysis results of the artifact removal model}
\begin{center}
\begin{tabular}{|c|c|c|}
\hline
\textbf{Metrics/Model} & \textbf{Unpruned} & \textbf{Pruned}\\
\hline
\textbf{Model size (MB)} & 1.75 & 0.54\\
\hline
\textbf{Accuracy (\%)} & 95.13 & 92.78\\
\hline
\textbf{Inference time (ms) (per mini batch)} & 42 & 29\\
\hline
\end{tabular}
\label{artanalysis}
\end{center}
\vspace{-3ex}
\end{table}
\subsection{Model parameters}
We have carried out analysis using statistical classifiers, MLP and CNN based models. In case of statistical classifier we have to extract the features first, then we give them as input to the classifier. But in MLP and CNN, which are end-to-end learning algorithms, we can directly feed the data into them. It can do the features extraction in it's layers. Our train-test split as 70:30. After extensive hyperparameter tuning, the parameters for the highest test accuracy are listed in table~\ref{mlpparameter}. Hyperparameters for MLP and CNN are given in table~\ref{cnnparameter}.\\
\begin{table}[htbp]
\caption{MLP model hyperparameters}
\begin{center}
\begin{tabular}{|c|c|}
\hline
\textbf{Hyperparameters} & \textbf{Value} \\
\hline
Epoch & 150\\
\hline
Optimizer & Adam\\
\hline
Learning rate & 0.001\\
\hline
Batch size & 64\\
\hline
Dropout & 0.4\\
\hline
\end{tabular}
\label{mlpparameter}
\end{center}
\vspace{-3ex}
\end{table}
\begin{table}[htbp]
\caption{CNN model hyperparameters}
\begin{center}
\begin{tabular}{|c|c|}
\hline
\textbf{Hyperparameters} & \textbf{Value} \\
\hline
Number of convolutional layers & 2\\
\hline
Number of fully connected layers & 2\\
\hline
Loss function & Binary cross entropy\\
\hline
Activation function & Softmax\\
\hline
Batch size & 64\\
\hline
Epoch & 150\\
\hline
Optimizer & Adam\\
\hline
Learning rate & 0.001\\
\hline
Kernel size & 3x1\\
\hline
\end{tabular}
\label{cnnparameter}
\end{center}
\vspace{-3ex}
\end{table}

In addition, the parameters for the statistical classifiers are:\\
Random forest : ({n\_estimators = 50}, {max\_depth = 5}).\\
SVM : ({C} = 1, {kernel} = rbf).
\subsection{Magnitude based weight pruning}
This compression algorithm creates a list of all weights of each layer by descending order and importance of the weights are determined by their magnitude. In this function we can tweak some hyperparameters, they are:
\begin{itemize}
    \item Sparsity
    \item Scheduling
\end{itemize}
The \textit{Sparsity} term can be correlated as a threshold value. By using this sparsity value, we can define how much nodes we want to prune or in other words, how dense we want our model will be. In our case 50\% sparsity is used.

The term  \textit{scheduling} is used to determine the starting step, the pruning frequency and the end step. By using this, we can allow the model to reach a certain accuracy level, then the pruning procedure will start and will start to continue as the frequency defined in the function. The end step indicates the point of final pruning action. 
The main difference between the ‘Dropout’ strategy and this pruning is that, in dropout we can’t control the node elimination process. It just prunes the models in random manner. On the contrary, in magnitude-based weight pruning action, all the pruning action can be designed by the user.  As a result, a more specific result with more controlled system can be obtained.
\subsection{Implementation details}
\begin{figure}[htbp]
\centering
\centerline{\includegraphics[height=4cm, width=\linewidth]{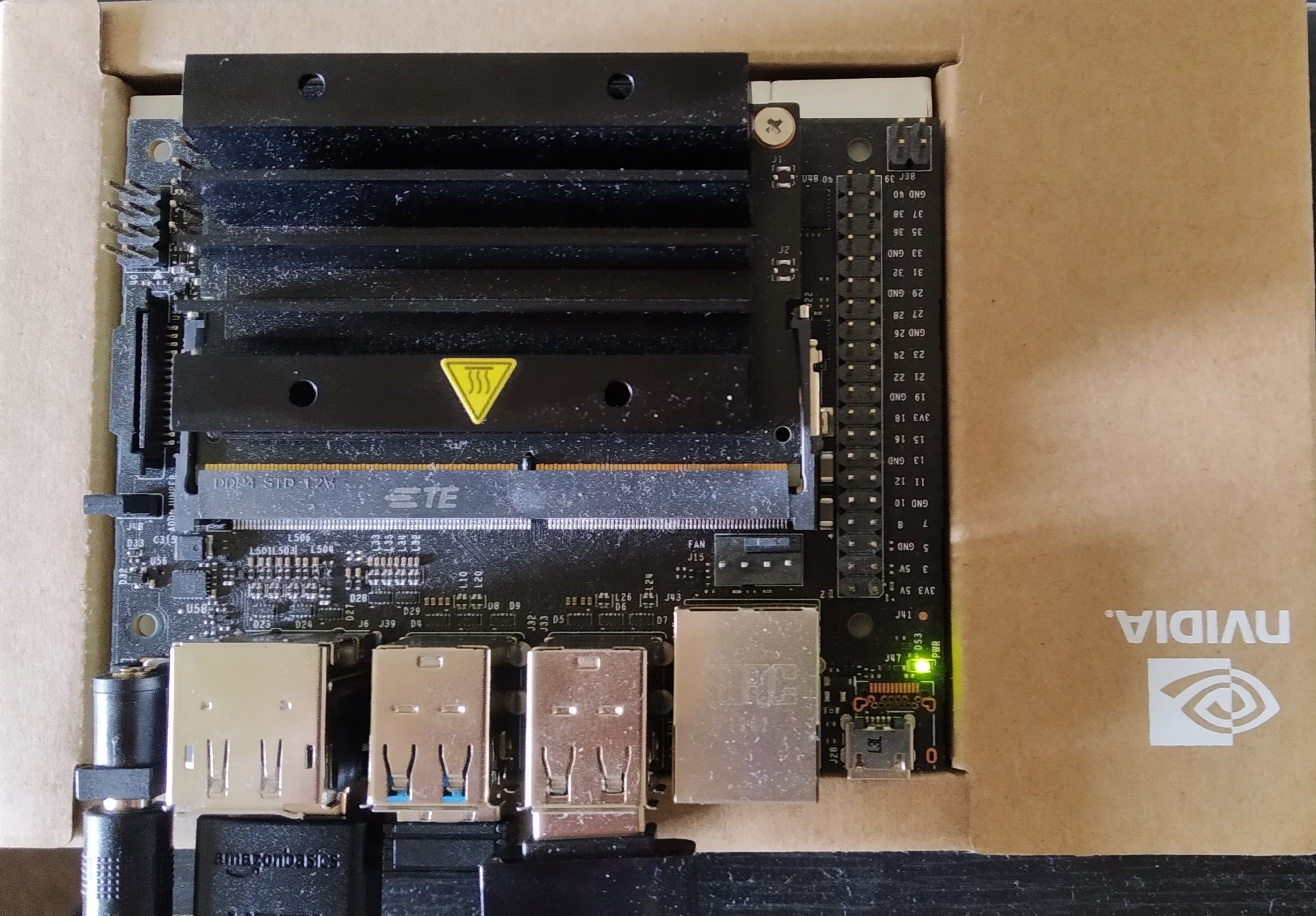}}
\caption{Experimental setup with Nvidia Jetson Nano.}
\label{nano}
\vspace{-1ex}
\end{figure}

The coding part of this work is carried out using Tensorflow (version 1.14) as deep learning library and Scikit-learn for implementation of statistical classifiers. For the hardware implementation task, as a resource constrained device, Nvidia Jetson Nano is used. It is one of the lowest configured based on CPU, Memory and perfectly suitable for this kind of IoT based development. In fig~\ref{nano} a snapshot of the hardware setup is provided. 
The device specifications are provided in table~\ref{specs}.
\begin{table}[htbp]
\caption{Nvidia Jetson Nano specifications}
\begin{center}
\begin{tabular}{|c|c|}
\hline
GPU	& 128-core Maxwell\\
\hline
CPU	& Quad-core ARM A57@1.43 GHz\\
\hline
Memory & 4 GB 64-bit LPDDR4 25.6 GB/s\\
\hline
\end{tabular}
\label{specs}
\end{center}
\vspace{-3ex}
\end{table}

\section{Evaluation}
We have divided our evaluation section into two major parts. In first part, we show the comparison of statistical models with uncompressed MLP model on the basis of accuracy and F1 score to understand the baseline performance with this dataset. Then the performance of compressed MLP and CNN models with respect to memory requirement and inference time will be compared in the second part.
\subsection{Comparison of statistical models with MLP}
The results for two statistical classifiers (Random Forest (RF) and SVM) and a 7 layer MLP model are shown in three different setups. First, we have used all the features and no artifact removal approach is applied. We can see the results in table~\ref{result}.

\begin{table*}[]
\caption{Experimental results comparison among three different models with three separate settings}
\begin{center}
\begin{tabular}{|c|c|c|c|c|c|c|c|c|c|}
\hline
\multirow{2}{*}{\textbf{\begin{tabular}[c]{@{}c@{}}Experimental settings/\\ Metrics\end{tabular}}} & \multicolumn{3}{c|}{\textbf{\begin{tabular}[c]{@{}c@{}}Using all features without \\ artifact removal\end{tabular}}} & \multicolumn{3}{c|}{\textbf{\begin{tabular}[c]{@{}c@{}}Using all features with \\ artifact removal\end{tabular}}} & \multicolumn{3}{c|}{\textbf{\begin{tabular}[c]{@{}c@{}}Using selected features with \\ artifact removal\end{tabular}}} \\ \cline{2-10} 
                                                                                                  & RF                                 & SVM                                & MLP (7 layers)                             & RF                                & SVM                               & MLP (7 layers)                            & RF                                  & SVM                                 & MLP (7 layers)                             \\ \hline
\textbf{Accuracy (\%)}                                                                            & 78.92                              & 68.53                              & \textbf{82.42}                                      & 78.87                             & 71.20                             & \textbf{84.88}                                     & \textbf{84.71}                               & 75.32                               & 79.04                                      \\ \hline
\textbf{F1 score}                                                                                 & 0.7977                             & 0.6318                             & \textbf{0.8015}                                     & 0.7732                            & 0.7001                            & \textbf{0.8229}                                    & \textbf{0.8114}                              & 0.7091                              & 0.7768                                     \\ \hline
\end{tabular}
\end{center}
\label{result}
\vspace{-4ex}
\end{table*}
If we closely analyze the result values for this setup, it is clear that MLP with 7 layers has the leverage. The model tends to do better with increasing number of layers but the relation is not linear. After going through experiments, this number of layers proved to be more stable. Random forest may not have the  highest accuracy but is also doing good.

Secondly, we have applied the artifact removal model output with all the features. The results are also given in table~\ref{result}.
Here, we can see the similar result as the previous one and the accuracy and F1 score have increased as a matter of fact. So, the effect of artifact removal is evident here.

In our third setting, we have fed the selected features after correlation check and t-test. The output are placed in table~\ref{result}. Surprising results can be noticed in the third setting. Random forest classifier surpasses the 7 layer MLP in this case. The reason behind this can be, as we have applied feature selection and artifact removal together, the statistical classifier gets a better scenario to make prediction and due to lack of some features MLP shows some less accuracy. It is acceptable because being a self feature extracting architecture, deep learning models is expected to do better with more number of features whereas statistical learning based classifiers has the upper hand with previously extracted and fewer number of features. Overall, the highest accuracy is achieved by all the features with artifacts removed. So, we can conclude saying that artifact removal is an important aspect and in case of deep learning the feature selection is less effective in our case.
\subsection{Evaluation of the compressed models}
\begin{figure}[htbp]
  \centering
  \subfloat []{\includegraphics[width=0.24\textwidth]{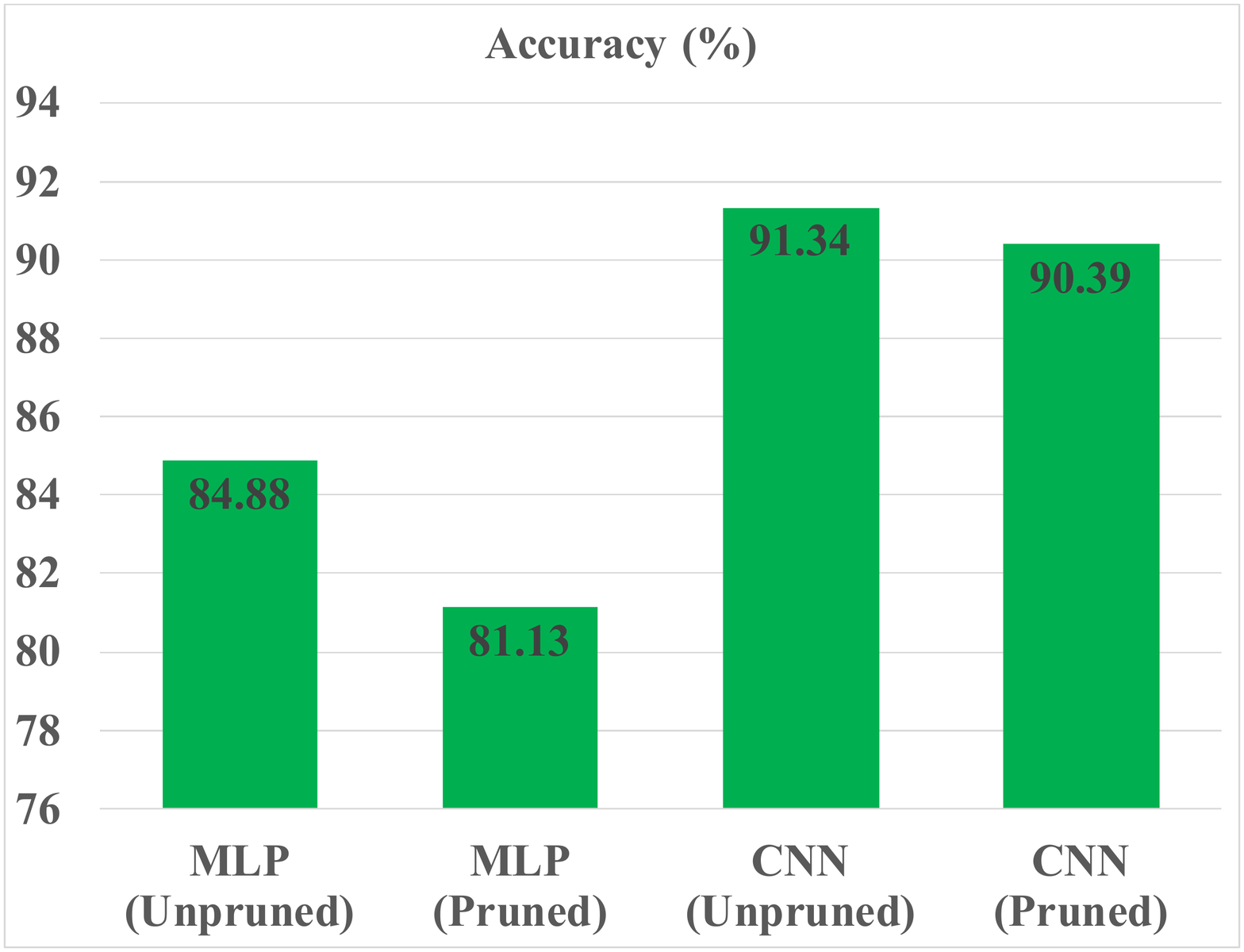}}
  \hfill
  \subfloat []{\includegraphics[width=0.24\textwidth]{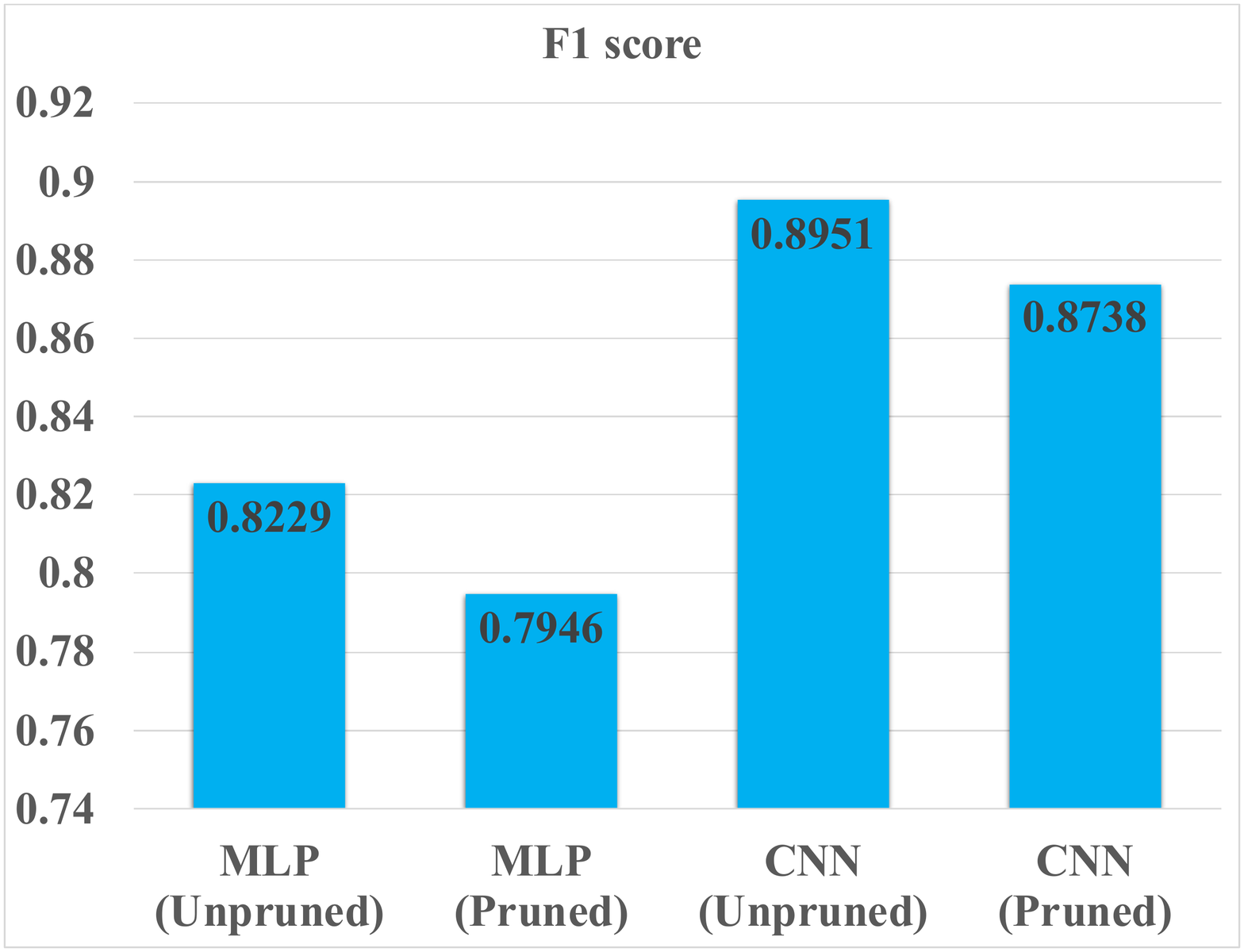}}
  \caption{(a) Accuracy and (b) F1 score comparison between pruned and unpruned models of MLP and CNN (with artifact removal).}
  \label{accuracy}
\vspace{-2ex}
\end{figure}

\begin{figure}[htbp]
  \centering
  \subfloat []{\includegraphics[width=0.24\textwidth]{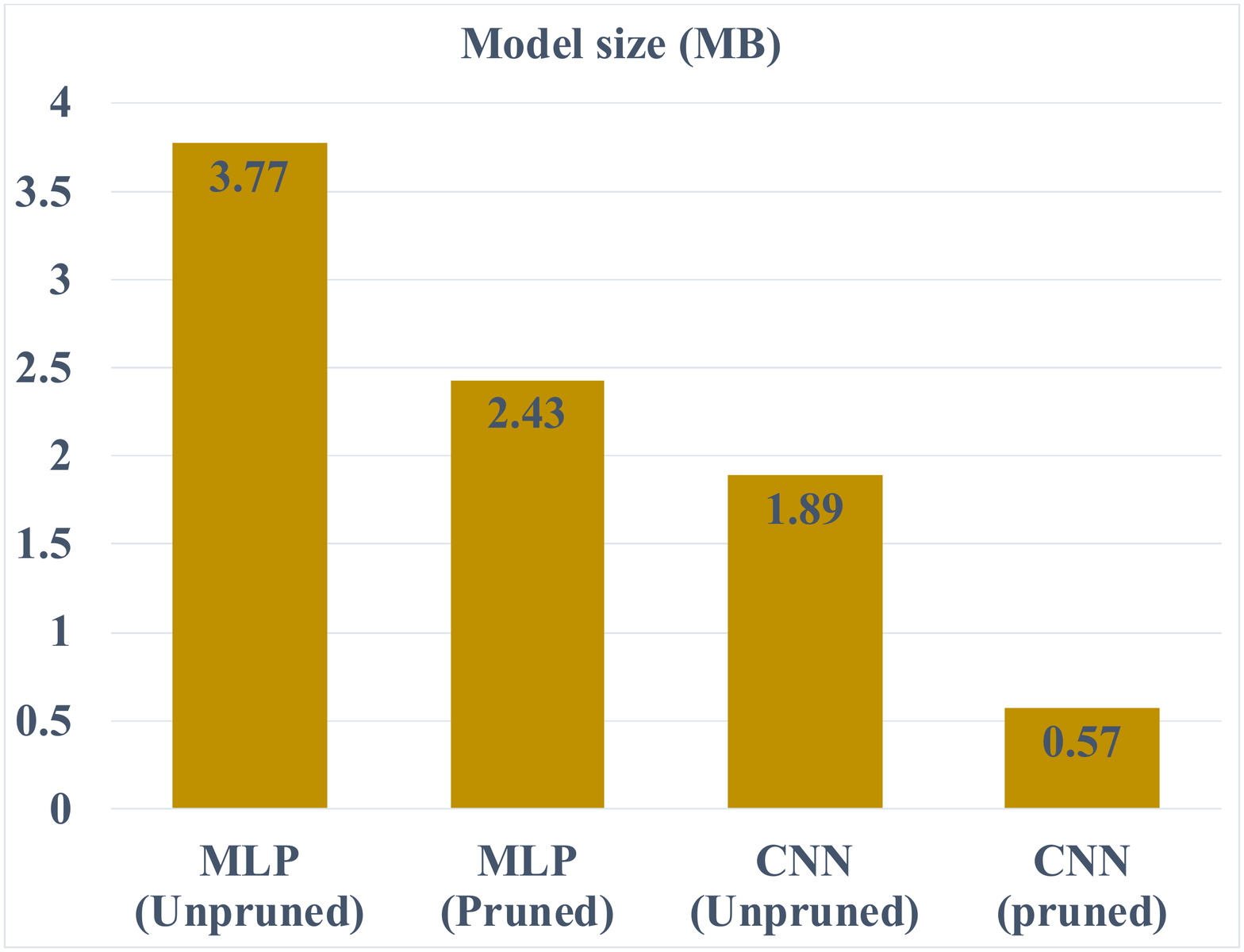}}
  \hfill
  \subfloat []{\includegraphics[width=0.24\textwidth]{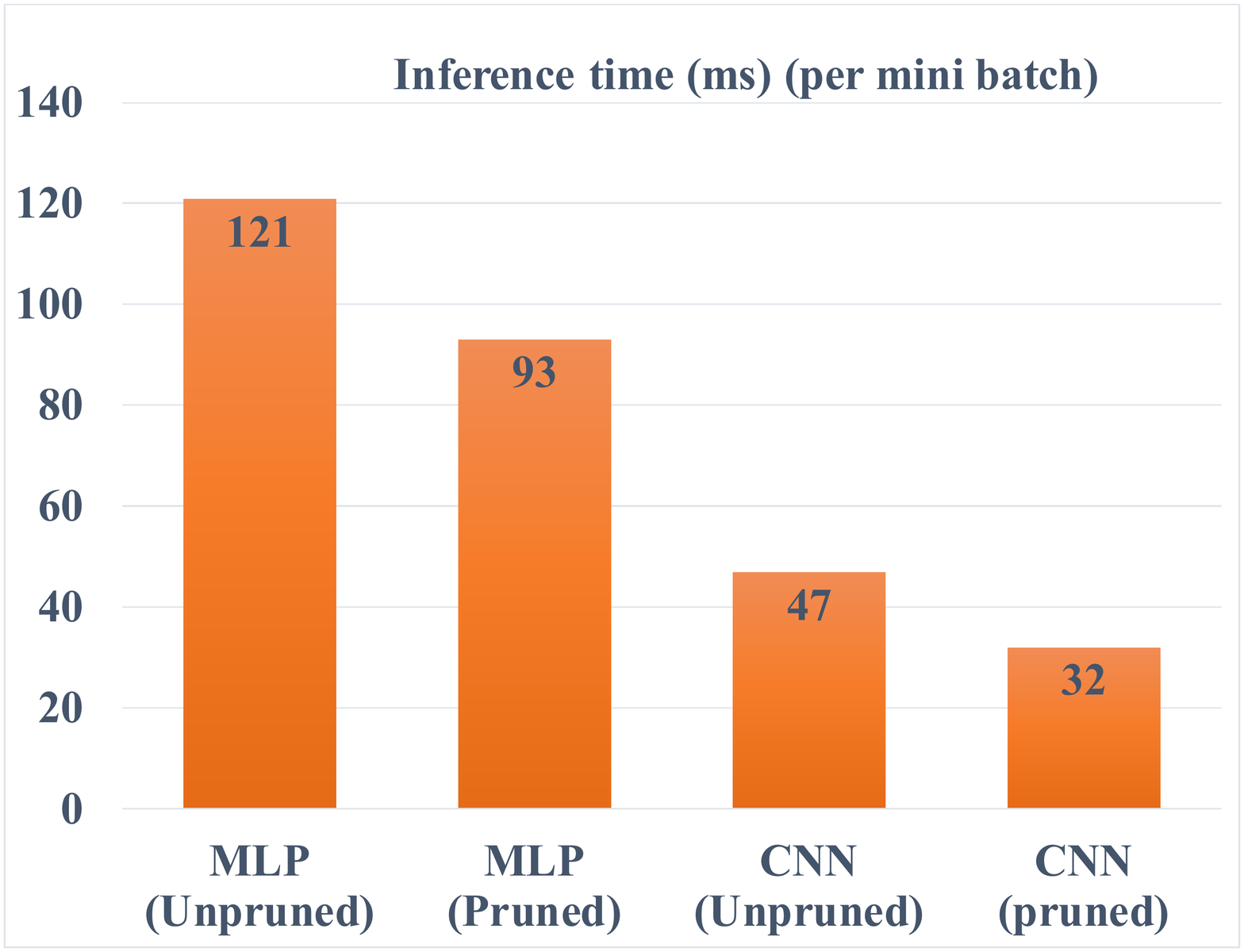}}
  \caption{(a) Model size (MB) and (b) inference time (ms) comparison between pruned and unpruned models of MLP and CNN.}
  \label{model size}
\vspace{-2ex}
\end{figure}

\begin{figure}[htbp]
\centering
\includegraphics[height=5cm, width=\linewidth]{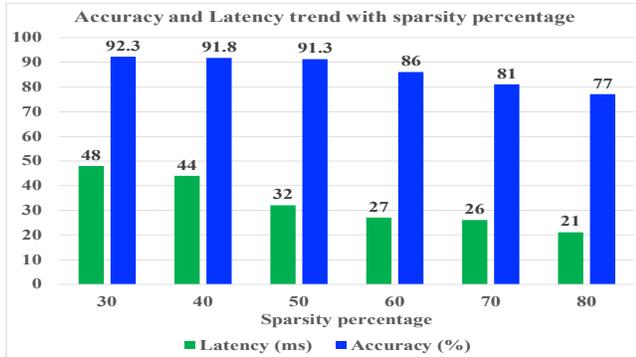}
\caption{Model size and latency trend with respect to sparsity percentage.}
\label{sparse}
\vspace{-2ex}
\end{figure}

The analysis results of both MLP (7 layer) and CNN are carried out on their pruned and unpruned forms. The values of model size and inference time are listed for every model. The comparison of those models based on accuracy and F1 score can be found in fig~\ref{accuracy}. It can be noticed that the accuracy drop is less than 3\% in all of the compressed models. In fig~\ref{model size}, we can also see a clear difference on the model size between the pruned and unpruned model. To be specific, in case on MLP there is a 35\% reduction in the model size which is 70\% in case of CNN. Significant improvement can also be seen in inference time also. A 23\% faster inference can be noticed in MLP and the value is 31\% in case of CNN.
An empirical analysis on the dependency of model size and latency on the percentage of the sparsity is depicted in fig~\ref{sparse}. Analyzing the trend and after going through some trade off between accuracy and latency, we have finalized 50\% sparsity in the best setting for our experiment. 
\section{Conclusion and Future Work}
In this work, we have demonstrated a EEG data based {\it on-device mental behaviour anomaly detection (OMAD)} model for substance and non-substance user groups. We have also deployed a light-weight MLP based artifact removal model trained on our own collected data. The main EEG data are fed to this model and the suspected artifact data points are removed. The higher testing accuracy (92.78\%) of the model ensures a better segregation between artifact and important brain signals.
From evaluation results, it is evident that artifact removal can help with more accurate results. From the analytic results of the compressed models, we can claim to achieve over 70\% reduction of model size with less than 3\% loss in accuracy for CNN. Also, we have managed to gain 31\% faster inference time for the same compressed CNN model. In future, we want to implement this model on EEG data of gaseous substance user and apply more sophisticated model compression techniques for analysis. Recent trends of segmenting deep model like distributed training, federated learning can also come up with interesting results. Also, we want to explore a signal processing driven approach to represent the analysis results based on alpha, beta, gamma and theta frequency components. 
\section*{Acknowledgment}
This research is partially supported by the NSF CAREER Award \# 1750936, ONR under grant N00014-18-1-2462, and Alzheimer’s Association, Grant/Award \# AARG-17-533039.

\bibliographystyle{unsrt}
\bibliography{main}

\begin{thebibliography}{10}

\bibitem{Longterm}
Effects of covid-19 long term.
\newblock
  \url{https://www.healthline.com/health-news/what-we-know-about-the-long-term-effects-of-covid-19#Who-is-most-at-risk?}
\newblock Accessed: 2020-05-18.

\bibitem{Samsha}
Drug abuse control.
\newblock \url{https://www.samhsa.gov/grants/grant-announcements/fg-20-006}.
\newblock Accessed: 2020-05-18.

\bibitem{AARP}
Effects on brain aarp.
\newblock
  \url{https://www.aarp.org/health/conditions-treatments/info-2020/covid-19-brain-symptoms.html}.
\newblock Accessed: 2020-05-18.

\bibitem{brain}
Effects on brain.
\newblock
  \url{https://www.wired.com/story/what-does-covid-19-do-to-your-brain/}.
\newblock Accessed: 2020-05-18.

\bibitem{nan2019deep}
Kaiming Nan, Sicong Liu, Junzhao Du, and Hui Liu.
\newblock Deep model compression for mobile platforms: A survey.
\newblock {\em Tsinghua Science and Technology}, 24(6):677--693, 2019.

\bibitem{tang2017enabling}
Jie Tang, Dawei Sun, Shaoshan Liu, and Jean-Luc Gaudiot.
\newblock Enabling deep learning on iot devices.
\newblock {\em Computer}, 50(10):92--96, 2017.

\bibitem{lane2017squeezing}
Nicholas~D Lane, Sourav Bhattacharya, Akhil Mathur, Petko Georgiev, Claudio
  Forlivesi, and Fahim Kawsar.
\newblock Squeezing deep learning into mobile and embedded devices.
\newblock {\em IEEE Pervasive Computing}, 16(3):82--88, 2017.

\bibitem{wu2018lightweight}
Tong Wu, Wenfeng Zhao, Edward Keefer, and Zhi Yang.
\newblock A lightweight deep compressive model for large-scale spike
  compression.
\newblock In {\em 2018 IEEE Biomedical Circuits and Systems Conference
  (BioCAS)}, pages 1--4. IEEE, 2018.

\bibitem{lane2015early}
Nicholas~D Lane, Sourav Bhattacharya, Petko Georgiev, Claudio Forlivesi, and
  Fahim Kawsar.
\newblock An early resource characterization of deep learning on wearables,
  smartphones and internet-of-things devices.
\newblock In {\em Proceedings of the 2015 international workshop on internet of
  things towards applications}, pages 7--12, 2015.

\bibitem{cohen2017does}
Michael~X Cohen.
\newblock Where does eeg come from and what does it mean?
\newblock {\em Trends in neurosciences}, 40(4):208--218, 2017.

\bibitem{jiang2019removal}
Xiao Jiang, Gui-Bin Bian, and Zean Tian.
\newblock Removal of artifacts from eeg signals: a review.
\newblock {\em Sensors}, 19(5):987, 2019.

\bibitem{wang2015changes}
Grace~Y Wang, Rob Kydd, Trecia~A Wouldes, Maree Jensen, and Bruce~R Russell.
\newblock Changes in resting eeg following methadone treatment in opiate
  addicts.
\newblock {\em Clinical Neurophysiology}, 126(5):943--950, 2015.

\bibitem{whitton1978spectral}
Joel~L Whitton, Frank Lue, and Harvey Moldofsky.
\newblock A spectral method for removing eye movement artifacts from the eeg.
\newblock {\em Electroencephalography and clinical neurophysiology},
  44(6):735--741, 1978.

\bibitem{anderer1999artifact}
Peter Anderer, Stephen Roberts, Alois Schl{\"o}gl, Georg Gruber, Gerhard
  Kl{\"o}sch, Werner Herrmann, Peter Rappelsberger, Oliver Filz, Manel~J
  Barbanoj, Georg Dorffner, et~al.
\newblock Artifact processing in computerized analysis of sleep eeg--a review.
\newblock {\em Neuropsychobiology}, 40(3):150--157, 1999.

\bibitem{rosler2013first}
Oliver R{\"o}sler and David Suendermann.
\newblock A first step towards eye state prediction using eeg.
\newblock {\em Proc. of the AIHLS}, 2013.

\bibitem{han2015learning}
Song Han, Jeff Pool, John Tran, and William Dally.
\newblock Learning both weights and connections for efficient neural network.
\newblock In {\em Advances in neural information processing systems}, pages
  1135--1143, 2015.

\bibitem{li2016pruning}
Hao Li, Asim Kadav, Igor Durdanovic, Hanan Samet, and Hans~Peter Graf.
\newblock Pruning filters for efficient convnets.
\newblock {\em arXiv preprint arXiv:1608.08710}, 2016.

\bibitem{wu2016quantized}
Jiaxiang Wu, Cong Leng, Yuhang Wang, Qinghao Hu, and Jian Cheng.
\newblock Quantized convolutional neural networks for mobile devices.
\newblock In {\em Proceedings of the IEEE Conference on Computer Vision and
  Pattern Recognition}, pages 4820--4828, 2016.

\bibitem{gong2014compressing}
Yunchao Gong, Liu Liu, Ming Yang, and Lubomir Bourdev.
\newblock Compressing deep convolutional networks using vector quantization.
\newblock {\em arXiv preprint arXiv:1412.6115}, 2014.

\bibitem{kim2015compression}
Yong-Deok Kim, Eunhyeok Park, Sungjoo Yoo, Taelim Choi, Lu~Yang, and Dongjun
  Shin.
\newblock Compression of deep convolutional neural networks for fast and low
  power mobile applications.
\newblock {\em arXiv preprint arXiv:1511.06530}, 2015.

\bibitem{mathur2018using}
Akhil Mathur, Tianlin Zhang, Sourav Bhattacharya, Petar Velickovic, Leonid
  Joffe, Nicholas~D Lane, Fahim Kawsar, and Pietro Li{\'o}.
\newblock Using deep data augmentation training to address software and
  hardware heterogeneities in wearable and smartphone sensing devices.
\newblock In {\em 2018 17th ACM/IEEE International Conference on Information
  Processing in Sensor Networks (IPSN)}, pages 200--211. IEEE, 2018.

\bibitem{hinton2015distilling}
Geoffrey Hinton, Oriol Vinyals, and Jeff Dean.
\newblock Distilling the knowledge in a neural network.
\newblock {\em arXiv preprint arXiv:1503.02531}, 2015.

\bibitem{zhang2015efficient}
Xiangyu Zhang, Jianhua Zou, Xiang Ming, Kaiming He, and Jian Sun.
\newblock Efficient and accurate approximations of nonlinear convolutional
  networks.
\newblock In {\em Proceedings of the IEEE Conference on Computer Vision and
  pattern Recognition}, pages 1984--1992, 2015.

\bibitem{snodgrass1980standardized}
Joan~G Snodgrass and Mary Vanderwart.
\newblock A standardized set of 260 pictures: norms for name agreement, image
  agreement, familiarity, and visual complexity.
\newblock {\em Journal of experimental psychology: Human learning and memory},
  6(2):174, 1980.

\end{thebibliography}
\end{document}